\documentclass{aa}  

\usepackage{graphicx}
\usepackage{lscape}
\usepackage{subfigure}
\usepackage{natbib}
\usepackage[varg]{txfonts}
\usepackage{comment}
\usepackage{diagbox}

\usepackage{longtable}
\usepackage{booktabs}
\usepackage[normalem]{ulem} 
\usepackage{epstopdf}
\usepackage{xcolor}
\usepackage{soul}
\usepackage{makecell}
\usepackage{multicol,tabularx,capt-of}
\usepackage{multirow}

\usepackage{scalerel}
\usepackage{tikz}

\usetikzlibrary{svg.path}

\defcitealias{Sanchez-Saez21a}{SS21a}

\definecolor{orcidlogocol}{HTML}{A6CE39}
\tikzset{
  orcidlogo/.pic={
    \fill[orcidlogocol] svg{M256,128c0,70.7-57.3,128-128,128C57.3,256,0,198.7,0,128C0,57.3,57.3,0,128,0C198.7,0,256,57.3,256,128z};
    \fill[white] svg{M86.3,186.2H70.9V79.1h15.4v48.4V186.2z}
                 svg{M108.9,79.1h41.6c39.6,0,57,28.3,57,53.6c0,27.5-21.5,53.6-56.8,53.6h-41.8V79.1z M124.3,172.4h24.5c34.9,0,42.9-26.5,42.9-39.7c0-21.5-13.7-39.7-43.7-39.7h-23.7V172.4z}
                 svg{M88.7,56.8c0,5.5-4.5,10.1-10.1,10.1c-5.6,0-10.1-4.6-10.1-10.1c0-5.6,4.5-10.1,10.1-10.1C84.2,46.7,88.7,51.3,88.7,56.8z};
  }
}

\newcommand\orcidicon[1]{\href{https://orcid.org/#1}{\mbox{\scalerel*{
\begin{tikzpicture}[yscale=-1,transform shape]
\pic{orcidlogo};
\end{tikzpicture}
}{|}}}}

\usepackage{hyperref} 
\hypersetup{
    colorlinks=true,
    citecolor=blue,
    linkcolor=blue,
    urlcolor=blue,
    }
\makeatletter
\renewcommand*\aa@pageof{, page \thepage{} of \pageref*{LastPage}}
\makeatother

\newcommand\rev[1]{#1}

\begin{document}

   \title{ATAT: Astronomical Transformer for time series And Tabular data}


   \author{
   G. Cabrera-Vives\inst{1,2,3\thanks{\email{guillecabrera@inf.udec.cl}}\orcidicon{0000-0002-2720-7218}},
   D. Moreno-Cartagena\inst{1,2},
   N. Astorga\inst{3,4,5},
   I. Reyes-Jainaga\inst{3,4},
   F. F\"orster\inst{6,3\orcidicon{0000-0003-3459-2270}},
   P. Huijse\inst{7,8,3\orcidicon{0000-0003-3541-1697}},
   J. Arredondo\inst{3,4\orcidicon{0000-0002-2045-7134}},
   A. M. Mu\~noz Arancibia\inst{3,4\orcidicon{0000-0002-8722-516X}},
   A. Bayo\inst{9,10\orcidicon{0000-0001-7868-7031}}, 
   M. Catelan\inst{11,3\orcidicon{0000-0001-6003-8877}},
   P. A. Est\'evez\inst{5,3\orcidicon{0000-0001-9164-4722}},
   P. S\'anchez-S\'aez\inst{9,3\orcidicon{0000-0003-0820-4692}},
   A. Álvarez\inst{3,4},
   P. Castellanos\inst{2},
   P. Gallardo\inst{2,3,4},
   A. Moya\inst{3,4},
   D. Rodriguez-Mancini\inst{12}
   }

\institute{
Department of Computer Science, Universidad de Concepci\'on, Chile
\and
Center for Data and Artificial Intelligence, Universidad de Concepción, Edmundo Larenas 310, Concepci\'on, Chile
\and
Millennium Institute of Astrophysics (MAS), Nuncio Monse\~nor Sotero Sanz 100, Of. 104, Providencia, Santiago, Chile
\and
Center for Mathematical Modeling (CMM), Universidad de Chile, Beauchef 851, Santiago 8320000, Chile
\and
Department of Electrical Engineering, Universidad de Chile, Av. Tupper 2007, Santiago 8320000, Chile
\and 
Data and Artificial Intelligence Initiative (ID\&IA), University of Chile, Santiago, Chile
\and
Institute of Astronomy (IvS), Department of Physics and Astronomy, KU Leuven, Celestijnenlaan 200D, 3001 Leuven, Belgium
\and
Instituto de Informática, Facultad de Ciencias de la Ingeniería, Universidad Austral de Chile, General Lagos 2086, Valdivia, Chile
\and
European Southern Observatory, Karl-Schwarzschild-Strasse 2, 85748 Garching bei München, Germany
\and
Instituto de F\'isica y Astronom\'ia, Universidad de Valpara\'iso, Av. Gran Breta\~na 1111, Playa Ancha, Casilla 5030, Chile
\and
Centro de Astroingenier{\'{\i}}a, Pontificia Universidad Cat{\'{o}}lica de Chile, Av. Vicu\~{n}a Mackenna 4860, 7820436 Macul, Santiago, Chile
\and
Data Observatory Foundation, Diagonal Las Torres N°2640, edificio E, Peñalolén, Santiago, Chile
}
   \date{Received September 15, 1996; accepted March 16, 1997}

 
  \abstract
   {The advent of next-generation survey instruments, such as the Vera C. Rubin Observatory and its Legacy Survey of Space and Time (LSST), is opening a window for new research in time-domain astronomy. 
   The Extended LSST Astronomical Time-Series Classification Challenge (ELAsTiCC) was created to test the capacity of brokers to deal with a simulated LSST stream.}
  {\rev{We aim to develop a next-generation model for classification of variable astronomical objects. We describe ATAT, the Astronomical Transformer for time series And Tabular data, a classification model conceived by the ALeRCE alert broker to classify light-curves from next-generation alert streams. ATAT was tested in production during the first round of the ELAsTiCC campaigns.}}
  {ATAT consists of two Transformer models that encode light curves and features using novel time modulation and quantile feature tokenizer mechanisms, respectively. ATAT was trained on different combinations of light curves, metadata, and features calculated over the light curves. \rev{We compare ATAT against the current ALeRCE classifier, a Balanced Hierarchical Random Forest (BHRF) trained on human-engineered features derived from light curves and metadata.}}
   {When trained on light curves and metadata, ATAT achieves a macro F1-score of $82.9\pm 0.4$ in 20 classes, outperforming the BHRF model trained on 429 features, which achieves a macro F1-score of $79.4\pm 0.1$.} 
   {The use of Transformer multimodal architectures, combining light curves and tabular data, opens new possibilities for classifying alerts from a new generation of large etendue telescopes, such as the Vera C. Rubin Observatory, in real-world brokering scenarios.}


    \titlerunning{ATAT: Astronomical Transformer for time series And Tabular data}
    \authorrunning{G. Cabrera-Vives et al.}
 
    \maketitle
%

\section{Introduction}
\label{sec:intro}

A new generation of synoptic telescopes are carrying out data-intensive observation campaigns. An emblematic example is the Vera C. Rubin Observatory and its Legacy Survey of Space and Time (LSST) \citep{ivezic2019lsst}. Starting in \rev{2025}, the Rubin Observatory will generate an average of 10 million alerts and \rev{$\sim$20 TB of raw data every night}. The massive data-stream of LSST is to be distributed to Community Brokers\footnote{\url{https://www.lsst.org/scientists/alert-brokers}} that will be in charge of ingesting, processing and serving the annotated alerts to the astronomical community. Collaboration between astronomers, computer scientists, statisticians and engineers is key to solve the rising astronomical big-data challenges \citep{borne2010astroinformatics, huijse2014computational}.

Automatic data processing based on Feature Engineering \citep[FE, e.g.,][]{FATS} and Machine learning (ML), including Deep Learning (DL) have been applied extensively in astronomical data applications, such as light-curve and image-based classification  \citep[e.g.,][]{dieleman2015, cabrera2016, deepHits, carrasco2021, russeil2022finding}, clustering \citep{ClusteringMackPavlos, ClusteringTransients}, physical parameter estimation \citep{forster2018, sanchez2021amortized, villar2022amortized}, and outlier detection \citep{nun2016ensemble, pruzhinskaya2019anomaly, sanchez2021, ishida2021active, perez-carrasco2023MCDSVDD, Perez-Carrasco_2023}. The Vera C. Rubin Community Brokers: ALeRCE \citep{ALERCE}, AMPEL \citep{2019A&A...631A.147N}, ANTARES \citep{matheson2021antares}, BABAMUL, Fink \citep{2021MNRAS.501.3272M}, Lasair \citep{2019RNAAS...3...26S}, and Pitt-Google\footnote{\url{https://pitt-broker.readthedocs.io/en/latest/}} are processing or will process massive amounts of data that is annotated with cross-matches, ML model predictions, and/or other information that is distributed to the community. These scientific products allow astronomers to study transient and variable objects in almost real-time or in an offline fashion for a systematic analysis of large numbers of objects. To enable the former, ML models should be integrated into a complex infrastructure and allow for accurate, rapid and scalable evaluation of tens of thousands of alerts received every minute \citep{2018ApJS..236....9N, rodriguez2022toward, cabrera2022managing}. In the past, the most common choices have been decision tree-based ensembles (e.g., Random Forest, RF, or Light Gradient Boosting Machine, LightGBM), models with high predictive performance, but high resource usage, due to the FE step. 
Several efforts have applied faster DL-based approaches to the problem of classifying astronomical time series (e.g., Recurrent Neural Networks, RNN, \cite{charnock2017deep, naul, carrasco2019, muthukrishna2019rapid, becker2020scalable, gomez2020, jamal2020neural, donoso2021}). These approaches usually focus on a specific type of object, e.g. variable stars \citep{naul, becker2020scalable, jamal2020neural} or transients \citep{charnock2017deep, muthukrishna2019rapid, Moller2019, fraga2024transient}. 
Until this work, ALeRCE had not been able to outperform tree-based ensembles \citep{Boone_2019, PLASTIC, neira2020mantra, Sanchez-Saez_2021, sanchez-saez2023persistent} with deep learning approaches for real-time alert classification across a broad taxonomy encompassing transient, stochastic, and periodic variable objects simultaneously.

More recently, Multi-Head Attention \citep[MHA;][]{attention_you_need} and Transformers have appeared as promising alternatives to time series encoders in astronomy \citep{other_attention, oscar_attention, donoso2022astromer, moreno2023positional}. These models are faster than RNNs since they have access to all the input simultaneously and not sequentially.
The above mentioned attention-based models have not explored training with multiple data sources (time series, metadata and human-engineered features) simultaneously.


The astronomical community has made great efforts to create realistic scenarios to test ML models \citep{PLASTIC}, but none of them have contemplated an end-to-end ML pipeline, \textit{i.e.} from the data ingestion to the ML model's outputs. The recent Extended LSST Astronomical Time-Series Classification Challenge (ELAsTiCC\footnote{\href{https://project.lsst.org/meetings/rubin2022/agenda/extended-lsst-astronomical-time-series-classification-challenge-elasticc}{ELAsTiCC Challenge, link 1}}$^{,}$\footnote{\href{https://portal.nersc.gov/cfs/lsst/DESC_TD_PUBLIC/ELASTICC/}{ ELAsTiCC Challenge, link 2}}, see Methods) has appeared as a unique opportunity to test broker's pipelines and ML models in production. ELAsTiCC is a challenge created by the Dark Energy Science Collaboration (DESC) that simulates LSST-like astronomical alerts with the goal of connecting the LSST project, brokers, and DESC by testing end-to-end pipelines in real-time.
To fulfill this  objective, ELAsTiCC started an official data stream on September 28th, 2022. Additionally, ELAsTiCC provided a dataset to train ML models.

\begin{figure*}[t]
\centering
    \includegraphics[width=0.9\textwidth]{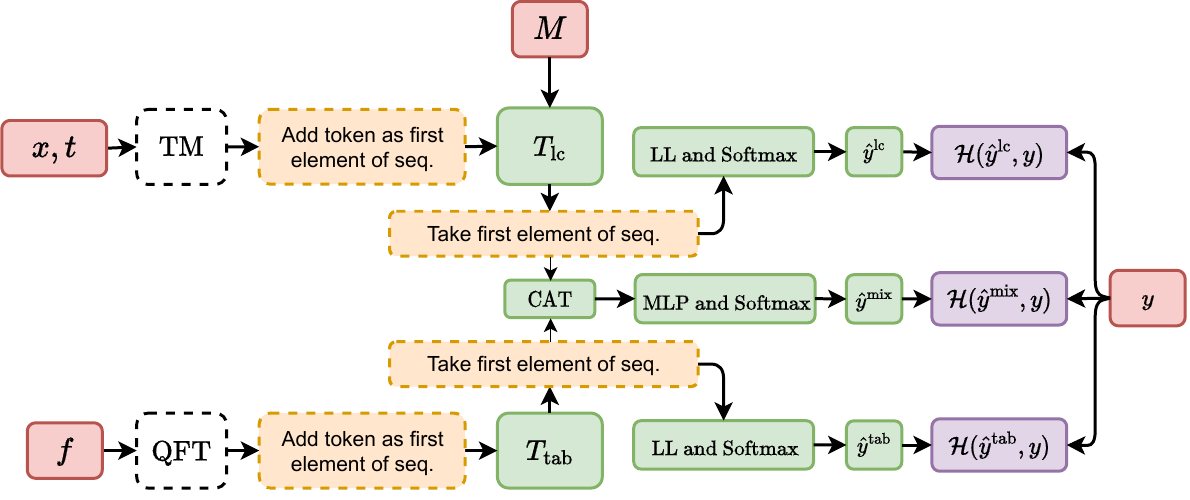}
      \caption{\label{atat} Diagram of ATAT, which consists of two branches: 1) on top a Transformer to process light-curves (matrices $x$, $t$ and $M$) and 2) at the bottom a Transformer to process tabular data (matrix $f$). Both information sources are processed by Time Modulation (TM) and Quantile Feature Tokenizer (QFT), respectively, represented as  
      white rectangles. In both cases, the results of this processing are sequences. Subsequently, a learnable token is added as the first element of the sequence. These sequences are processed by the Transformer architectures $T_{\text{lc}}$ (light-curves) and $T_{\text{tab}}$ (tabular data). Finally, the processed token is transformed linearly and used for label prediction ($\hat{y}^{\text{lc}}$ or $\hat{y}^{\text{tab}}$). In training we use cross-entropy $\mathcal{H}(\cdot, y)$ to optimize the model (purple rectangle). If both light-curves and tabular information are used at the same time, we additionally minimize the cross-entropy of the prediction  $\hat{y}^{\text{mix}}$ resulted from the concatenation of both processed tokens. In the diagram MLP, LL and CAT refers to Multi-Layer Perceptron, Linear Layer and concatenation in the embedding dimension, respectively. For more details see Methods.
      }\label{fig:ATAT}
\end{figure*}

In this work, we present the methods that ALeRCE used for the first round of ELAsTiCC. We propose ATAT, an Astronomical Transformer for time series And Tabular data, a model that is based on a Transformer architecture. ATAT is trained with the dataset provided by ELAsTiCC previous to the start of the real-time infrastructure challenge  and implemented as an end-to-end pipeline within the ALeRCE \citep{ALERCE} broker. ATAT can use time series information and all the available metadata and/or features obtained from other pre-processing steps (see Figure~\ref{fig:ATAT}). In summary, our contributions are:

\begin{itemize}
    \item A new state-of-the-art Transformer model called ATAT, which encodes multivariate, variable length, and irregularly-sampled light-curves in combination with metadata and/or extracted features. 
    \item A thorough comparison between ATAT and a RF-based baseline (historically the most competitive model of ALeRCE when applied to real data streams) in the ELAsTiCC dataset.
\end{itemize}

The code for replicating our experiments will be made publicly available once this article is accepted \footnote{\url{https://github.com/alercebroker/ATAT}}. Our code uses the publicly available ELAsTiCC dataset\footnote{\href{https://portal.nersc.gov/cfs/lsst/DESC_TD_PUBLIC/ELASTICC/TRAINING_SAMPLES/}{ELAsTiCC dataset}}.

\section{ELAsTiCC Overview and Machine Learning Approaches}

\subsection{ELAsTiCC} \label{elasticc}
\begin{figure*}[t]
\includegraphics[width=0.95\textwidth]{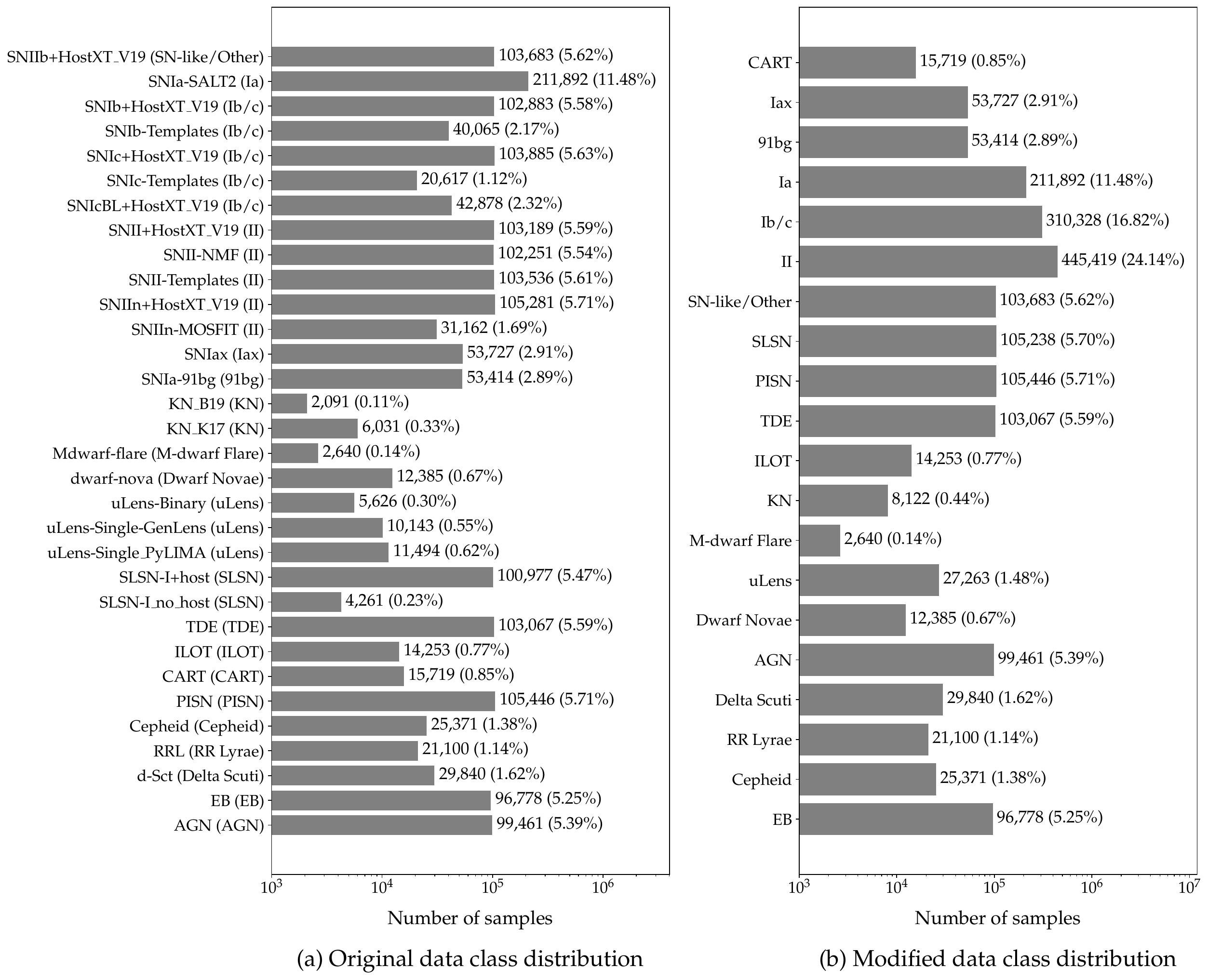}
      \caption{\label{histogram} ELAsTiCC dataset class histogram. In (a) the original taxonomy class distribution is shown. In (b) the taxonomy class distribution selected by Vera Rubin's brokers is shown. Note that we use the SN-like/Other class to include SNe IIb.}
\end{figure*}

The ELAsTiCC dataset contains 1,845,146 light-curves in six bands (ugrizY) from simulated astronomical objects distributed in 32 classes as shown in Figure \ref{histogram} (a). \rev{This work uses the training dataset from the first ELAsTiCC campaign.} We use the same taxonomy \rev{as} the ELAsTiCC broker's taxonomy, except for the SN-like/other class that includes only SNe IIb (see  Figure \ref{histogram} b). 

\rev{We consider 64 attributes from the metadata provided in the alert stream of ELAsTiCC. These attributes include the best heliocentric redshift, Milky Way extinction between the blue and visual (and its error), and for the first and second galaxy host match: ellipticity, magnitudes on each band (and their error), transient-host separation, radius, photometric redshift (and its error), deciles of the estimated photometric redshift probability density function, and spectroscopic redshift if available (and its error).}

We split the ELAsTiCC dataset into training, validation, and test sets. The test set contains one-thousand \rev{light-curves} of each class, ensuring a balanced representation. The remaining data is further stratified into five folds, from which we train five separate models (same folds for all neural networks and random forest). In each training iteration, one fold serves as a validation set for hyperparameter tuning and early stopping.
All the metrics presented in this paper are derived from evaluations performed on the test set.

Additionally, the ELAsTiCC dataset was modified by discarding information that is not available in the ELAsTiCC alert stream. For this purpose, we use the PHOTFLAG\footnote{\href{https://portal.nersc.gov/cfs/lsst/DESC_TD_PUBLIC/ELASTICC/TRAINING_SAMPLES/A_FORMAT.TXT}{Specific format information for the ELAsTiCC dataset.}} key to select only non-saturated data. 
\rev{We consider alerts and forced photometry points in each light-curve spanning from thirty days prior to the initial alert (forced photometry data for the first alert) to the final detection. If there are non-detections after the last detection, these are not included.}

\subsection{ATAT}

\label{ATAT}

Here we describe our proposed transformed-based model, ATAT, and the techniques developed to process time series information (light-curves) and tabular data information (metadata and/or processed features). For the rest of the paper, we will call these models ATAT's variants since different input combinations can be used.

For each astronomical source, we consider two types of data: the light-curve and tabular data composed of static \emph{metadata}  (e.g., host galaxy redshifts, if any) and features calculated from the light-curves (e.g., the period of a periodic source). For a particular source, an observation $j$ in band $b$ of its light-curve is described by the observation time $t_{j, b}$ and by the photometric data $x_{j, b} =(\mu_{j, b}, \sigma_{j, b})$, where $\mu_{j, b}$ represents the difference flux\footnote{Source flux density measured from a difference image.}, and $\sigma_{j, b}$  the flux error. Not all light-curves have the same number of observations. In order to represent this in the model input, we consider fixed size light-curves of the length of the largest light-curve in the dataset\rev{, consisting of 65 observations,} and perform zero padding (add zeros for observations after the maximum time in each band, both for $x_{j, b}$ and $t_{j, b}$). At the same time, not all bands are observed \rev{simultaneously}. This is represented by adding zeros to $\mu_{j, b}$ and $\sigma_{j, b}$ of the unobserved bands of observation $j$. In order to mask attention for these unobserved values, we use a binary mask $M_{j, b}$ such that $M_{j, b} = 1$ if observation $j$ is observed at band $b$, and $M_{j, b} = 0$ if not \citep{attention_you_need, bert}. For each source, tabular data consists of $K$ features $f_k$, $k \in \{1, \dots, K\}$ which, as explained above, may be \rev{metadata or features calculated from the light-curves}.

As a first step, time series and tabular data are processed using Time Modulation and a Quantile Feature Tokenizer, respectively. These steps return sequences that can be used as inputs for common Transformer architectures. Figure~\ref{atat} shows a general scheme of ATAT. Its hyperparameters are further specified in Implementation details section. 
We noted larger models showed better performance, but we limited their size to reduce the memory requirements in production. For the rest of the paper we will denote a linear layer as $\text{LL}$. 

\subsubsection{Time Modulation}
\label{TM-sec}
Time Modulation (TM) incorporates time information of observation $j$ and band $b$, $t_{j,b}$, into the difference flux $\mu_{j,b}$ and flux error $\sigma_{j,b}$. Previous works have successfully applied TM in attention models, using processes similar to positional encoding \citep{attention_you_need}
. We construct a variant of the time modulation proposed by \cite{oscar_attention}, which is based on a Fourier decomposition. For each observation $j$ and band $b$ of the light-curve we perform a linear transformation on the input vector, transforming $x_{j, b} = (\mu_{j,b}, \sigma_{j,b})$ to a vector $\mathrm{LL_{TM}}(x_{j, b})$ of dimension $E_\mathrm{TM}$\footnote{\rev{$\mathrm{LL_{TM}}(x)=W_\mathrm{TM}x$, where $W_\mathrm{TM}$ is a $E_\mathrm{TM}\times 2$ matrix.}}. We modulate this vector by doing an element-wise product with the output of a vector function $\gamma_b^1(t_{j, b})$ and add the output of a second vector function $\gamma_b^{2}(t_{j, b})$:
\begin{equation}
    \text{TM}(x_{j, b}, t_{j, b}) =  \text{LL}_{\text{TM}}(x_{j, b}) \odot \gamma_b^1(t_{j, b}) + \gamma_b^2(t_{j, b}).
    \label{tm}
\end{equation}

We define the functions $ \gamma_b^1(t_{j, b})$ and $ \gamma_b^2(t_{j, b})$ as Fourier series 
\begin{equation}
\begin{split}
    \gamma_b^1(t) = \sum_{h = 1}^H \alpha_{b,h}^1 \sin \left(\frac{2\pi h}{T_{\mathrm{max}}} t \right) + \beta_{b,h}^1  \cos \left(\frac{2\pi h}{T_{\mathrm{max}}} t \right),\\
    \gamma_b^2(t) = \sum_{h = 1}^H \alpha_{b,h}^2 \sin \left(\frac{2\pi h}{T_{\mathrm{max}}} t \right) + \beta_{b,h}^2  \cos \left(\frac{2\pi h}{T_{\mathrm{max}}} t \right),
    \label{fourier}
\end{split}
\end{equation}
where $T_{\mathrm{max}}$ is an hyperparameter that is set higher than the maximum timespan of the longest light-curve in the dataset, $H$ is the number of harmonics in the Fourier series, and $\alpha_{b,h}^1, \beta_{b,h}^1, \alpha_{b,h}^2$, and $\beta_{b,h}^2$ are learnable Fourier coefficients. \rev{In this work, $t$ is the number of days since the first forced photometry point in the light curve.}

Eq.~\eqref{tm} applies a linear transformation to $x_{j, b}$, expanding its dimension. After that, a scale and bias are created as Fourier series (Eq.~\ref{fourier}) using time $t_{j, b}$. Note that a Fourier series can have enough expressive power for large $H$. Eq.~\eqref{tm} is applied separately for each band, and their output vectors are later concatenated in the sequence dimension. Consequently, the output of TM for a light-curve is a matrix of dimension $L \cdot B \times E_{\text{TM}}$, where $B$ is the number of bands, and $L$ is the maximum number of observations that a band can have for all bands and light-curves in the dataset.
 
\subsubsection{Quantile Feature Tokenizer}
\label{sec:QFT}

Tabular data in ELAsTiCC may include processed light-curve features, static metadata, or a concatenation of both. We process this data before feeding it into a Transformer. We call this process Quantile Feature Tokenizer (QFT) and it comprises two steps. First, a quantile transformation\footnote{A quantile transformation transforms features into a desired distribution \rev{(a normal distribution in this work)} by mapping the cumulative distribution function of the features to the quantile function of the desired distribution.} $\mathrm{QT}_k(f_k)$ is applied to each feature $f_k$ of the tabular data of each object, to normalize them as a way to deal with complex distributions \rev{(e.g. skewed) and help the predictive model perform better}. Second, an affine transformation is used to vectorize each scalar value of the attributes recorded in the tabular data
\begin{equation}
    \text{QFT}_k(f_{k}) = W_k \cdot \text{QT}_k(f_{k}) + b_k,
    \label{tt}
\end{equation}
where $\cdot$ stands for matrix multiplication, $k$ refers to the index feature, and $W_k$ and $b_k$ are vectors of learnable parameters with dimensions $E_{\text{QFT}}$. In other words, the $k^{th}
$ scalar feature $f_k$ is transformed by $\text{QT}_k$ and then vectorized by multiplying it by $W_k$ and adding $b_k$. Notice that a different transformation is applied to each feature of the tabular data. The output dimension $E_{\text{QFT}}$ is an hyperparameter to be chosen. This methodology is similar to \cite{FeatureTokenizer}, but we additionally apply the quantile transformation to each feature that is fitted before training the model.

\subsubsection{Transformers}
\label{Transformer-sec}

The Transformer architecture is based on Bidirectional Encoder Representations from Transformers \citep[BERT,][]{bert} and Vision Transformers \citep{VIT} which aim at processing sequential information. The architecture consists of a multi-head attention (MHA) step and a forward fully-connected (FF) neural network step with skip connections.

The transformer architecture used in this work can be described as follows: consider a layer $l$ which receives as input the output of layer $l-1$. Then, our transformer can be described as
\begin{align}
h_*^l &= \text{MHA}^{l-1}(h^{l-1}) + h^{l-1}, \label{mha} \\
h^l &= \text{FF}^{l-1}(h_*^l) + h_*^l \label{FF},
\end{align}
where $l \in [1 \dots n^{layers}]$ ($n^{layers}$ being the number of layers of the Transformer), $h^l$ is the output of layer $l$, and $h^l_*$ is the output of the MHA step which includes a skip connection. Notice $h_*^l$ serves as input to a feed-forward network with a skip connection which outputs $h^l$. Relevant Transformer hyperparameters include the number of heads $n^{heads}$ and the embedding dimensionality $E_{T}$, which are specified in Section \ref{sec:ImplementationDetails}. Additionally, we use a learnable classification token of dimension $E_{T}$ that is concatenated at the beginning of the input sequence. This token representation after the Transformer is fed into an output layer that performs the classification task. The dimension number of $E_{T}$ is equal to $E_{\text{TM}}$ for the light-curve Transformer  and $E_{\text{QFT}}$  for the tabular data Transformer (equal to their input). 

When only a single data source is considered (e.g., only light-curves data), we take the first element of the Transformer's output sequence, and apply a linear layer plus a softmax activation function. When two data sources are considered, two Transformers $T_{\text{lc}}$ and $T_{\text{tab}}$ are used to process light-curve and tabular data information, respectively. The first output elements of both sequences are concatenated, and a multilayer perceptron plus a softmax activation function are applied to produce the label prediction (see Figure~\ref{atat}).

\subsubsection{Mask temporal augmentation}
\label{sec:MTA}

To improve early classification performance, we train ATAT on light-curves reduced up to a randomly selected time instant. During training, a day $t^* \in \{8, 128, 2048\}$ is randomly selected for each light-curve, and the values of mask $M$ corresponding to times $t>t^*$ are set to zero. Note that $t^*=2048$ is equivalent to using the complete light-curves. Times are selected from a limited discrete set since it is unfeasible to compute the features at arbitrary times. Hereafter, we refer to this augmentation method as Masked Temporal Augmentation (MTA). Considering the lengths of the light-curves during training is a standard procedure that has been used in the past \citep{Moller2019, donoso2021, Gagliano_2023}.

\subsubsection{Implementation details}
\label{sec:ImplementationDetails}

ATAT variants are evaluated every twenty-thousand iterations, and early stopping with a patience of three evaluations is used. Models are trained using the Adam optimizer \cite{Adam} until early stopping with learning rate of $2 \cdot 10^{-4}$ and a batch size of 256. We use class balanced batches. \rev{The fully connected neural network of Eq. \ref{FF} consists of two layers. The first layer employs a Gaussian Error Linear Unit  \citep[GELU;][]{hendrycks2016gaussian} while the second layer is a linear layer.} For both $T_{\text{lc}}$ and $T_{\text{tab}}$, $n^{heads}=4$ and $n^{layers}= 3$. For $T_{\text{lc}}/T_{\text{tab}}$ all input and output dimensions of linear layers are 48/\rev{32} with the exception of the hidden layers of FF (Eq. \ref{FF}) which are 96/\rev{64}. Note that this implies that $E_{\text{TM}} = 48 \cdot 4 = 192$ and $E_{\text{TT}} = 36 \cdot 4 = 144$. We select $T_{max} = 1500$ \rev{days} and $H = 64$. We use a dropout of 0.2 for training. All Nans, $\inf$ and $-\inf$ in features and metadata are replaced by -9999.

\subsection{Features}

The ELAsTiCC dataset has six bands and its light-curves contain difference fluxes. In comparison, the alert stream from ZTF \citep{bellm2018, graham2019} classified by the original \rev{Balanced Hierarchical Random Forest model from ALeRCE} has only two fully-public bands and it offers light-curves in difference magnitudes. In order to deal with this, we modified some of the original features from \cite{Sanchez-Saez_2021}. 
All light-curve based features were modified to use fluxes as input instead of magnitudes.
The supernova parametric model (SPM) from \cite{Sanchez-Saez_2021} was modified to better handle the six bands available and the extra information of redshift and Milky Way dust extinction. The fluxes were scaled using the redshift information available and the WMAP5 cosmological model \citep{komatsu2009, 2022ApJ...935..167A}, and also deattenuated using the extinction information and the model from \cite{o1994rnu}. This means that some of metadata information was used in the computation of features. We remove some features from \cite{Sanchez-Saez_2021} that were not simulated by ELAsTiCC, e.g., the star-galaxy score from the ZTF stream and the color information from ALLWISE. The coordinates of the objects are not used because they were not simulated in a realistic way for each of the astrophysical classes. We ended up with a total of 429 features: 69 features per band plus 15 multi-band (e.g. colors, multi-band periodogram, among others). These engineered features are also used for ATAT in Section \ref{results-sec}. In Appendix \ref{processedfeat} we give a comprehensive list of the modified features. 

\rev{We calculate features for all light-curves with a total of more than 5 points across all bands. Since we are using forced photometry, the source with fewer observations has 11 points. Consequently, we calculate features for all objects.}

\subsection{Balanced Hierarchical Random Forest}
\label{bhrf}

We compare our Transformer models against the Balanced Hierarchical Random Forest (BHRF) model of \cite{Sanchez-Saez_2021} adapted for the ELAsTiCC dataset. This section describes the differences between the original BHRF and its ELAsTiCC adaptation.

The original BHRF described in \cite{Sanchez-Saez_2021} is composed of four balanced random forest models \citep{Chen2004} that are used in a hierarchical structure. The top model classifies each light-curve into Transient, Stochastic and Periodic classes. Then each one of these three groups is further classified using its own Balanced Random Forest model. In this work, the Transient group includes the following classes: Calcium Rich Transients (CART), SNe Iax (Iax), SNe 91bg-like (91bg), SNe Ia (Ia), SNe Ib/c (Ib/c), SNe II (II), SNe IIb (SN-like/Other), Superluminous SNe (SLSN), Pair Instability SNe (PISN), Tidal Disruption Events (TDE), Intermediate Luminosity Optical Transients (ILOT), and Kilonovae (KN). The Stochastic group includes the following classes: M-dwarf flare, Dwarf novae, active galactic nuclei (AGN), and gravitational micro lensing events (uLens). The Periodic group includes the following classes: Delta Scuti, RR Lyrae, Cepheid, and Eclipsing Binary.

\rev{During training, the balanced random forest models draw bootstrap samples from the minority class and sample with replacement the same number of samples from the other classes for each tree. In order to be consistent with ATAT, the balanced random forest models were trained using light-curves trimmed to 8, 128 and 2048 days long. We sampled 15,000, 9,000, and 3,900 light-curves from each class to train each tree of the transient, periodic, and stochastic balanced random forests, respectively. Notice that all data was fed to the model during training, but each tree only uses a randomly sampled balanced subset of the data.}
We performed a hyperparameter grid search for the split criterion (Gini impurity, entropy), number of trees (10, 100, 350, 500), and maximum depth (10, 100, and no maximum depth). The best combination of hyperparameters in terms of the mean macro F1-score \rev{over predictions for the full light-curves} were obtained for an entropy split criterion, 350 trees, and a maximum depth of 100. We also calculated the feature importance on the random forest and used only 100 features for each forest which lead to a statistically similar macro F1-score (within one standard deviation). 
These decisions were made to diminish the final size of the model in order to facilitate deployment.

\section{Results}
\label{results-sec}

\begin{figure*}[ht]
\centering
    \includegraphics[height = 200pt]{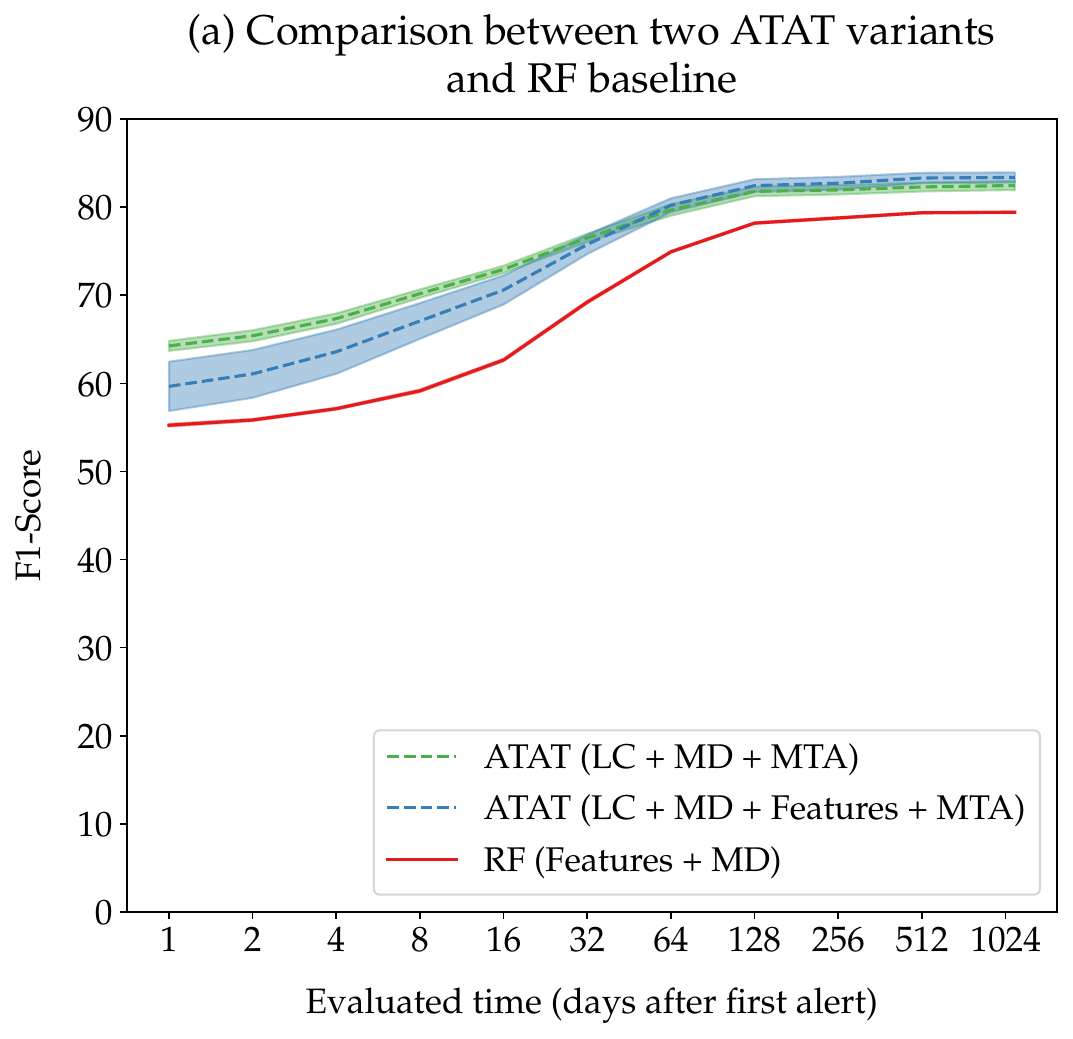}
    \includegraphics[height = 200pt]{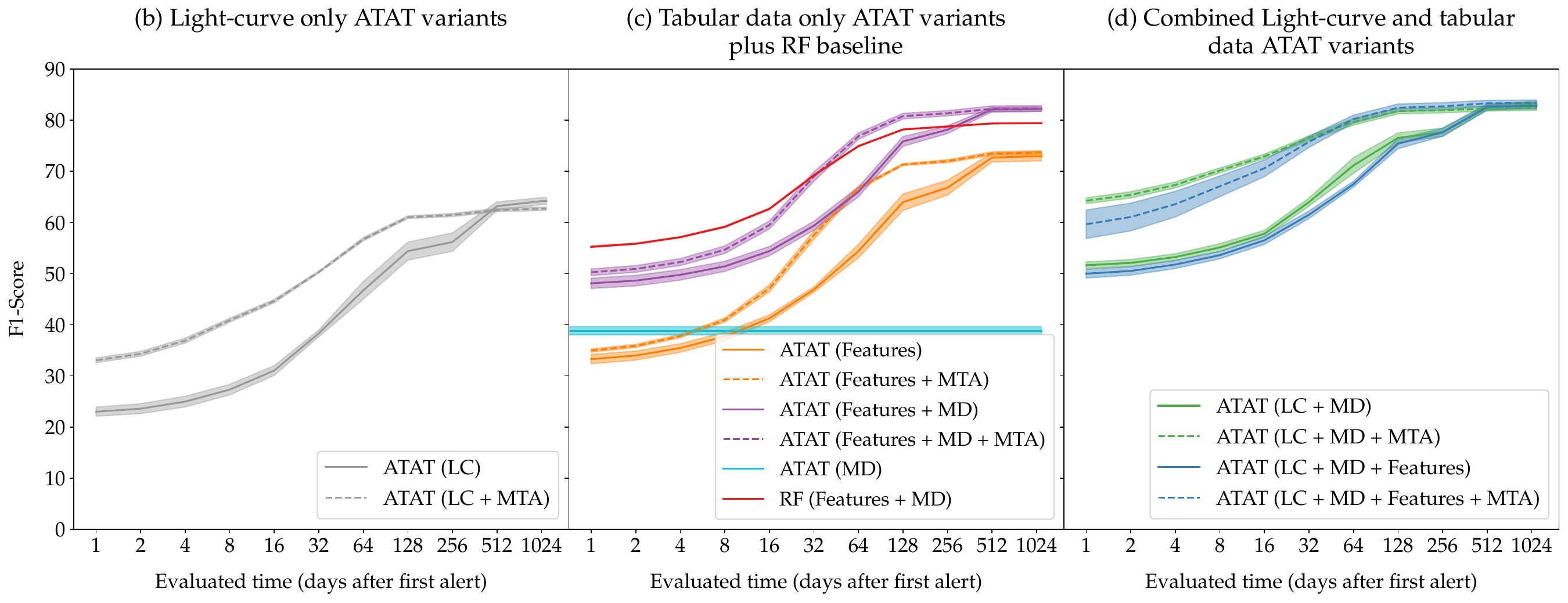}
      \caption{\label{fig:teval} F1-score vs time since first alert for a selection of models. We show the better performing ATAT variants and the RF-based baseline (a), the light-curve only ATAT variants (b), the tabular data only ATAT variants (c), and the combined light-curve and tabular data ATAT variants (d). LC/MD/Features refers to models that are optimized using the light-curve, metadata and feature information, respectively. Models can use more than one information source, e.g., LC + MD + Features.  Dotted lines refer to models that are optimized with MTA (see \rev{Section \ref{sec:MTA}).} }
\end{figure*}

\subsection{Comparison between ATAT and BHRF}

\rev{We calculate measures of classification performance by averaging predictions for each object.} Figure~\ref{fig:teval} (a) shows the test-set F1-score of two selected ATAT variants using different data sources with MTA and the BHRF as a function of the number of days after the first alert. \rev{The analysis begins on the day of the first detection and progressively increases in powers of two until reaching the maximum duration of the longest light-curve, which spans 1,104 days.}
These ATAT variants outperform the BHRF model for all light-curve lengths, specially for shorter light-curves. We further compare these models by measuring the F1-score, recall and precision in a per-class basis as shown in Table \ref{atattable}. The labels indicate whether the models were trained using the light-curve (LC) data, metadata (MD), engineered features (Features), or combinations of these. MD is the data coming together with the alerts (e.g., host galaxies redshift), while features are extracted from the light curves (e.g., period). 
The ATAT's variants surpass the BHRF in 16 out of 20 classes. In particular, the ATAT variant based on LC and MD performs better in the SN subclasses, but the ATAT variants that use all data sources obtain better scores in the periodic sub-classes. Three of the four classes where the BHRF outperforms (F1-Score) ATAT, namely KN, CART and M-dwarf flares are also the ones with fewer examples in the dataset (see Figure~\ref{histogram}). This may be explained by the differences in the class-balancing strategies, with the BHRF being more robust to overfitting in the minority classes. The BHRF model also outperforms ATAT in terms of the F1-Score for the Eclipsing Binaries (EBs), primarily because  ATAT achieves a relatively low precision for this class. As illustrated in Figure~\ref{CF}, about 10\% of the M-dwarf flares are classified as EBs by ATAT, which can be explained by the low support of M-dwarf flares in our dataset.

\subsection{Classification performance of ATAT variants}

\begin{table*}
\footnotesize
    \centering
\resizebox{\textwidth}{!}{
\begin{tabular}{cccccccccc}
\toprule 
Classnames & 
\multicolumn{3}{c}{ATAT (LC + MD)} & 
\multicolumn{3}{c}{ATAT (LC + MD +  Features)}&
\multicolumn{3}{c}{RF (MD +  Features)}\\
\cmidrule(r){2-4}
\cmidrule(r){5-7}
\cmidrule(r){8-10}
 & Precision & Recall & F1-Score
 & Precision & Recall & F1-Score
 & Precision & Recall & F1-Score \\
\midrule 
  CART & $72.8 \pm 2.9$ & $40.3 \pm 3.8$ & $51.7 \pm 2.9$ & \textbf{75.3 $\pm$ 2.5} & $40.0 \pm 4.6$ & $52.0 \pm 3.3$ & $59.2 \pm 0.4$ &  \textbf{56.2 $\pm$ 0.6} & \textbf{57.6 $\pm$ 0.5}\\
  Iax & \textbf{59.8 $\pm$ 1.3} & \textbf{70.1 $\pm$ 2.2} & \textbf{64.5 $\pm$ 1.2} & \textbf{59.8 $\pm$ 2.1} & $65.1 \pm 5.4$ & $62.2 \pm 1.9$ & $57.6 \pm 0.5$ &  $55.9 \pm 0.6$ &  $56.8 \pm 0.5$ \\
  91bg & \textbf{89.4 $\pm$ 0.4} & $92.4 \pm 1.1$ & \textbf{90.9 $\pm$ 0.4} & $88.8 \pm 2.2$ & \textbf{92.5 $\pm$ 1.9} & $90.5 \pm 0.6$ & $75.2 \pm 0.4$  & $90.2 \pm 0.2$  & $82.0 \pm 0.2$\\
  Ia & $75.7 \pm 1.6$ & \textbf{84.0 $\pm$ 1.7} & \textbf{79.6 $\pm$ 0.6} & \textbf{76.3 $\pm$ 1.2} & $81.4 \pm 1.7$ & $78.8 \pm 0.7$ & $61.4 \pm 0.4$ &  $76.7 \pm 0.2$ & $68.2 \pm 0.3$\\
  Ib/c & $53.8 \pm 3.1$ & $63.9 \pm 1.9$ & \textbf{58.3 $\pm$ 1.3} & $50.0 \pm 3.8$ & \textbf{65.8 $\pm$ 3.4} & $56.6 \pm 1.2$ & \textbf{58.0 $\pm$ 0.3} & $39.6 \pm 0.4$  & $47.1 \pm 0.2$ \\
  II & $65.8 \pm 2.3$ & $65.2 \pm 1.0$ & \textbf{65.5 $\pm$ 1.3} & $63.9 \pm 3.5$ & \textbf{66.4 $\pm$ 2.8} & $65.0 \pm 1.3$ & \textbf{66.8 $\pm$ 0.6} & $42.7 \pm 0.5$ & $52.1 \pm 0.5$ \\
  SN-like/Other & \textbf{66.7 $\pm$ 1.5} & \textbf{62.9 $\pm$ 2.4} & \textbf{64.7 $\pm$ 1.7} & $64.3 \pm 2.2$ & $60.5 \pm 2.9$ & $62.3 \pm 1.5$ & $59.0 \pm 0.5$ &  $54.1 \pm 0.8$ &  $56.5 \pm 0.6$ \\
  SLSN & $89.0 \pm 1.1$ & $95.3 \pm 0.3$ & $92.0 \pm 0.5$ & $89.6 \pm 0.9$ & \textbf{95.4 $\pm$ 0.4} & \textbf{92.4 $\pm$ 0.4} & \textbf{90.3 $\pm$ 0.1} & $90.0 \pm 0.1$ & $90.2 \pm 0.1$ \\
  PISN & $93.1 \pm 1.1$ & \textbf{97.5 $\pm$ 0.5} & $95.2 \pm 0.4$ & \textbf{95.9 $\pm$ 0.4} & $96.7 \pm 0.9$ & \textbf{96.3 $\pm$ 0.4} & $85.6 \pm 0.1$ & $96.7 \pm 0.1$  & $90.8 \pm 0.0$\\
  TDE & $77.2 \pm 2.9$ & \textbf{92.7 $\pm$ 0.8} & $84.2 \pm 1.6$ & $79.0 \pm 4.9$ & $92.5 \pm 1.0$ & \textbf{85.1 $\pm$ 2.6} & \textbf{83.2 $\pm$ 0.4} & $76.8 \pm 0.3$ &  $79.9 \pm 0.2$ \\
  ILOT & $89.6 \pm 0.8$ & $85.9 \pm 2.9$ & $87.7 \pm 1.1$ & \textbf{92.1 $\pm$ 0.9} & $84.0 \pm 3.1$ & \textbf{87.8 $\pm$ 1.3} & $76.3 \pm 0.3$ & \textbf{93.6 $\pm$ 0.2} & $84.1 \pm 0.2$\\
  KN & \textbf{97.7 $\pm$ 0.3} & $71.5 \pm 2.6$ & $82.6 \pm 1.7$ & $97.1 \pm 0.4$ & $77.1 \pm 2.5$ & $85.9 \pm 1.4$ &  $86.8 \pm 0.2$ & \textbf{90.3 $\pm$ 0.1} & \textbf{88.5 $\pm$ 0.1} \\
  M-dwarf Flare & $98.9 \pm 0.4$ & $67.5 \pm 1.2$ & $80.2 \pm 0.8$ & \textbf{99.1 $\pm$ 0.3} & $70.4 \pm 1.9$ & $82.3 \pm 1.3$ &  $95.0 \pm 0.3$ & \textbf{79.4 $\pm$ 0.3}  & \textbf{86.5 $\pm$ 0.3} \\
  uLens & $85.8 \pm 1.8$ & $95.3 \pm 0.9$ & $90.3 \pm 0.8$ & $86.8 \pm 1.7$ & \textbf{95.6 $\pm$ 0.7} & \textbf{91.0 $\pm$ 0.7} & \textbf{96.9 $\pm$ 0.4} & $82.8 \pm 0.2$ & $89.3 \pm 0.3$ \\
  Dwarf Novae & \textbf{87.9 $\pm$ 1.8} & $86.3 \pm 0.9$ & $87.1 \pm 0.7$ & $86.2 \pm 1.8$ & \textbf{92.0 $\pm$ 0.9} & \textbf{89.0 $\pm$ 0.9} & $78.5 \pm 0.2$ & $82.9 \pm 0.3$ &  $80.6 \pm 0.2$ \\
  AGN & \textbf{99.8 $\pm$ 0.0} & \textbf{100.0 $\pm$ 0.0} & \textbf{99.9 $\pm$ 0.0} & $99.7 \pm 0.1$ & \textbf{100.0 $\pm$ 0.0} & $99.8 \pm 0.1$ & $95.4 \pm 0.4$ & \textbf{99.9 $\pm$ 0.1} &  $97.6 \pm 0.2$ \\
  Delta Scuti & $92.3 \pm 0.7$ & $95.0 \pm 0.4$ & $93.6 \pm 0.3$ & \textbf{98.7 $\pm$ 0.3} & \textbf{99.5 $\pm$ 0.1} & \textbf{99.1 $\pm$ 0.1} & $90.8 \pm 0.3$ & $98.9 \pm 0.0$ &  $94.7 \pm 0.2$\\
  RR Lyrae & $93.5 \pm 1.0$ & $96.0 \pm 0.9$ & $94.7 \pm 0.3$ & \textbf{99.5 $\pm$ 0.2} & \textbf{99.1 $\pm$ 0.2} & \textbf{99.3 $\pm$ 0.1} & $91.6 \pm 0.4$ & $98.9 \pm 0.1$ & $95.1 \pm 0.2$ \\
  Cepheid & $96.0 \pm 0.8$ & $97.9 \pm 0.3$ & $97.0 \pm 0.4$ & \textbf{99.2 $\pm$ 0.3} & \textbf{99.5 $\pm$ 0.1} & \textbf{99.3 $\pm$ 0.1} & $92.6 \pm 0.5$ & $98.9 \pm 0.1$ & $95.6 \pm 0.3$ \\
  EB & $87.0 \pm 0.3$ & $98.8 \pm 0.3$ & $92.5 \pm 0.1$ & $90.4 \pm 1.7$ & \textbf{99.6 $\pm$ 0.1} & $94.8 \pm 0.9$ & \textbf{93.5 $\pm$ 0.3} & $97.5 \pm 0.1$ & \textbf{95.5 $\pm$ 0.2}\\
  \midrule
  Macro avg & $83.6 \pm 0.4$ & $82.9 \pm 0.5$ & $82.6 \pm 0.5$ & \textbf{84.6 $\pm$ 0.3} & \textbf{83.7 $\pm$ 0.6} & \textbf{83.5 $\pm$ 0.6} & $79.7 \pm 0.1$  & $80.1 \pm 0.1$ & $79.4 \pm 0.1$\\
\bottomrule 
\end{tabular}}
\vspace{.1cm}
    \caption{\label{atattable} Classification precision, recall and F1-score per class and macro average of the models put into production. We include the two best-performing ATAT variants (with MTA) and the RF-based baseline.}
    \label{tab:my_label}
\end{table*}

To explore the influence of the number and type of data sources on the classification performance, eleven ATAT variants are compared in Figures \ref{fig:teval} (b), (c) and (d). In all figures the dashed lines correspond to the cases where the MTA strategy is used. 

Figure \ref{fig:teval} (b) shows the performance of the ATAT variants using only the light-curve as input source, with and without the MTA strategy.  The MTA strategy significantly improves the classifier performance at early times, saturating at about 128 days and after that having only marginal increments. This could be related with the majority of the classes in the dataset being transients and with the absence of longer timescales variable objects (e.g., Miras and other LPVs). 

Figure \ref{fig:teval} (c) shows the performance of the ATAT variants trained using only tabular data information. This includes metadata from the first alert and features that are a function of the available light-curve data, where a strategy similar to MTA can also be applied. Note that the feature-based model (orange line) outperforms the LC-based model (grey line in Figure \ref{fig:teval} b).
\rev{We can also observe that the performance of the feature-based model increases considerably when metadata is incorporated (purple line), and even outperforms the RF-based baseline (red line) when considering MTA after $\sim$32 days. Before 32 days, the RF model outperforms all other ATAT models that only use tabular data. The ATAT model that uses only MD (cyan line) outperforms the ATAT that use only features for light-curves shorter than $\sim$8 days.} The MTA strategy applied to feature computation has a positive effect in early classification performance in all cases.

Figure~\ref{fig:teval} (d) shows the performance of four ATAT variants trained with both the light-curves and the metadata, with and without features, and with and without the MTA strategy. This figure suggests a low synergy between the light-curve and feature data and that ATAT can extract the most relevant class information using only the light-curve and metadata. Moreover, a comparison with Figure~\ref{fig:teval} (b) suggests a high synergy between the light-curve and metadata, where adding metadata yields a performance improvement between 20\% and 30\% depending on the length of light-curves. The ATAT variant that uses all information sources (blue solid-line) without the MTA strategy has a worse performance for light-curves shorter than 128 days than the model that uses only light-curve and metadata information (green solid-line). \rev{When the MTA strategy is applied, the model that uses all the data sources is only marginally superior after $\sim$32 days. Before $\sim$ 32 days, not using features leads to a higher macro F1-score.}

Summarizing, models combining light-curves and metadata information yield the highest performance (highest synergy). 
\rev{The addition of features to ATAT marginally improves the macro F1-score when calculated over the entire light-curves when using the MTA strategy.}
Additionally, applying the MTA strategy is always beneficial for early classification in the models using light-curves and/or feature data. The two ATAT variants that use light-curves plus metadata and MTA shown in Figure~\ref{fig:teval} (a) were put into production within the pipeline that processes the ELAsTiCC stream.

\begin{figure*}[t]
\centering
    \includegraphics[height = 280pt]{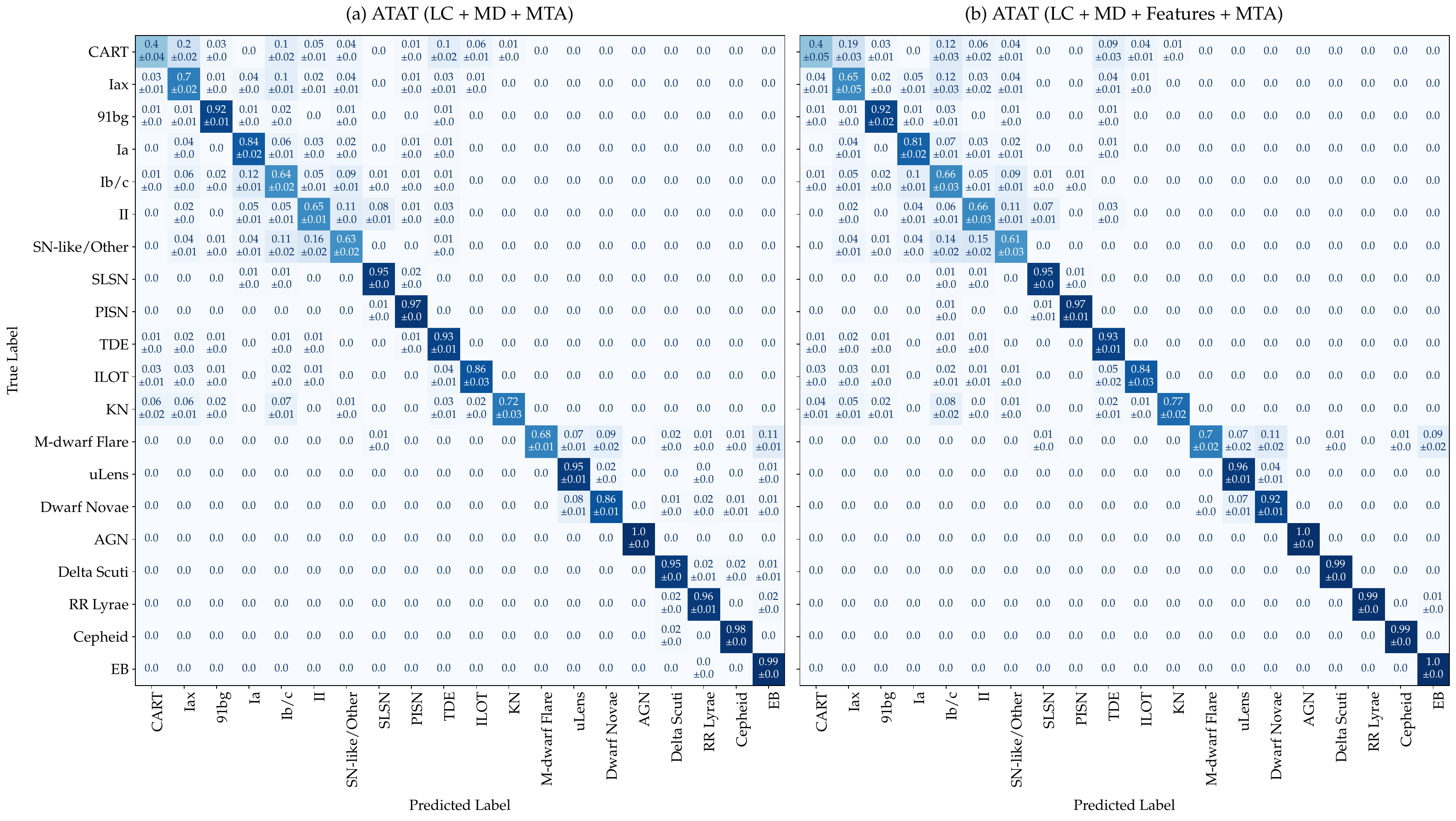}
    \includegraphics[height = 280pt]{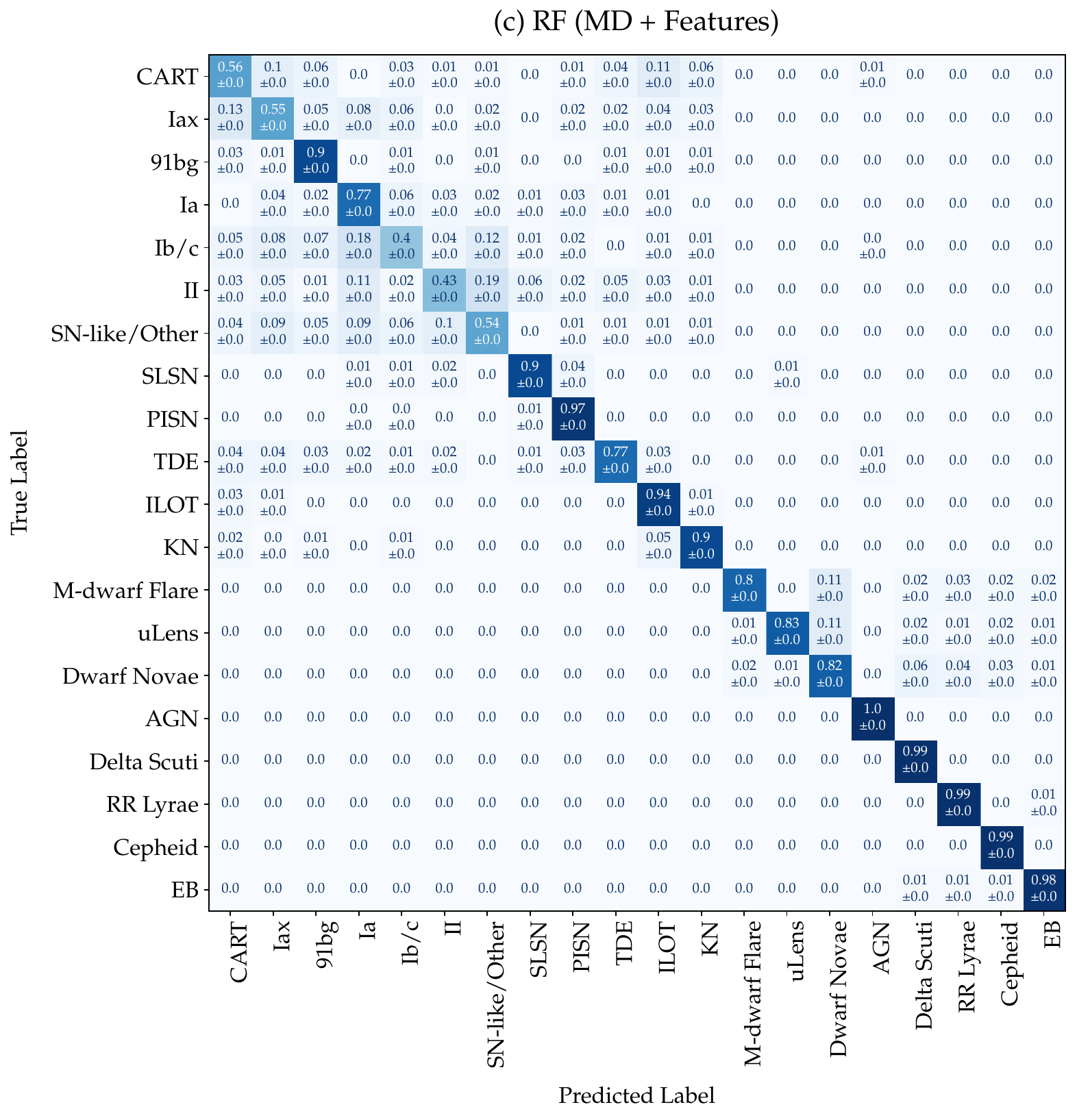}
      \caption{\label{CF} Confusion matrix of two ATAT variants, and the Random Forest (RF) baseline. ATAT (LC + MD + MTA) has an F1-Score of $82.6\%$, ATAT (LC + MD + Features + MTA) has an F1-Score of $83.5\%$, and RF (MD + Features) has an F1-Score of $79.4\%$.}
\end{figure*}

Figure~\ref{CF} shows the confusion matrices of: (a) the ATAT variant that uses LC, metadata and MTA; (b) the ATAT variant that uses LC, features, metadata, and MTA; and (c) the RF-based baseline. These results where obtained by evaluating the light-curves at their maximum length. The ATAT (LC + MD + Features + MTA) model outperformed the RF in 15 out of 20 classes in the dataset. In particular the ATAT performs better in all the SNe subclasses, namely: Iax, 91bg, Ia, Ib/c, II, SLSN, PISN and SN-like/Other. This is specially noticeable for types Ib/c and II where the difference in recall is 26\% and 23\%, respectively. In the case of transient types, besides the aforementioned SNe subclasses, noticeable differences between the models arise. For example the RF-based baseline outperforms ATAT by 13\% and 16\% in the case of KN and CART, respectively. The former model confused these classes mainly with SNe types Ib/c and Iax. The baseline is also 10\% better at detecting ILOT. ATAT confuses this class mainly with TDEs, whereas the baseline does not present such confusion. On the other hand, ATAT outperforms the baseline by 16\% and 13\% in the case of TDE and uLens, respectively. Cataclysmic types also present interesting differences between models. For example the ATAT outperforms the baseline by 10\% for the Dwarf-novae class. On the other hand, the RF-based baseline outperforms the ATAT model by 10\% in the case of M-dwarf flares. The latter model has 9\% confusion between this class and the EB periodic subtype, whereas the baseline does not present such confusion. Finally, the confusion matrices show that both models achieve almost perfect detection for periodic variable star classes and for the stochastic AGN class, with the proposed model being marginally superior than the baseline. It is worth noting that in three of the four classes where the RF-based baseline outperforms ATAT, namely KN, CART and M-dwarf are also the ones with fewer examples in the dataset (see Figure~\ref{histogram}). This suggests that the RF is more efficient for highly class-imbalanced datasets than the Transformer-based approach. Additional data-augmentation strategies may be required to improve the
performance of ATAT in these data-scarce classes. The ATAT variant that does not use features (LC + MD + MTA) shows a similar performance than the ATAT variant that uses them (LC + MD + Features + MTA), except for periodic classes where the feature-based variant is consistently better.

\subsection{Ablation study}

In this Section, we test the different design choices introduced in ATAT. Table~\ref{tab:ablation} shows the F1-Scores (mean and standard deviation) for combinations of positional encodings (PE) and tokenizers used in the feature transformer for the metadata. For the PE we consider our proposed time modulation (TM), the original fixed sinusoidal \cite{attention_you_need}, and not using the light curves at all (hence, no light curve transformer). For the tabular data transformer we consider our proposed Quantile Feature Tokenizer (QFT), the original feature tokenizer from Gorishniy et al. \cite{FeatureTokenizer}, and using no tabular metadata. For the tabular data we  consider the metadata and not the features, given that there is not more than 2$\sigma$ difference in the F1-Score with and without the features as shown in Figure~\ref{fig:teval}. All experiments include the MTA described in Section \ref{sec:MTA}. The combination of the TM and the QFT outperform other experiments. By using our proposed TM we increase the F1-Scores significantly (over 5\%) as compared to the sinusoidal PE, while our proposed QFT is able to increase the performance of our model by over 2\%.

\begin{table*}[ht]
\centering
\begin{tabular}{l|ccc}
\toprule
\multirow{2}{*}{\diagbox{Tab. tokenizer}{PE}} & \multirow{2}{*}{no lightcurve} & \multirow{2}{*}{fixed sinusoidal \cite{attention_you_need}} & time modulation \\
                  &                                &                                   & (this work)     \\
\midrule
No tabular data   & -                              & 52.56 $\pm$ 0.62                      & 62.70 $\pm$ 0.38    \\
FT                & 36.10 $\pm$ 0.28                   & 73.67 $\pm$ 0.33                      & 79.99 $\pm$ 0.63    \\
QFT (this work)              & 39.44 $\pm$ 0.50                   & 76.77 $\pm$ 0.45    & 82.61 $\pm$ 0.59    \\     
\bottomrule
\end{tabular}
\vspace{.1cm}
\caption{\label{tab:ablation} Ablation study F1-Scores. Each row represents a different feature tokenizer used for the tabular transformer. Each column represents a different positional encoder.}
\end{table*}

\begin{figure}[t]
\centering
    \includegraphics[width=.45\textwidth]{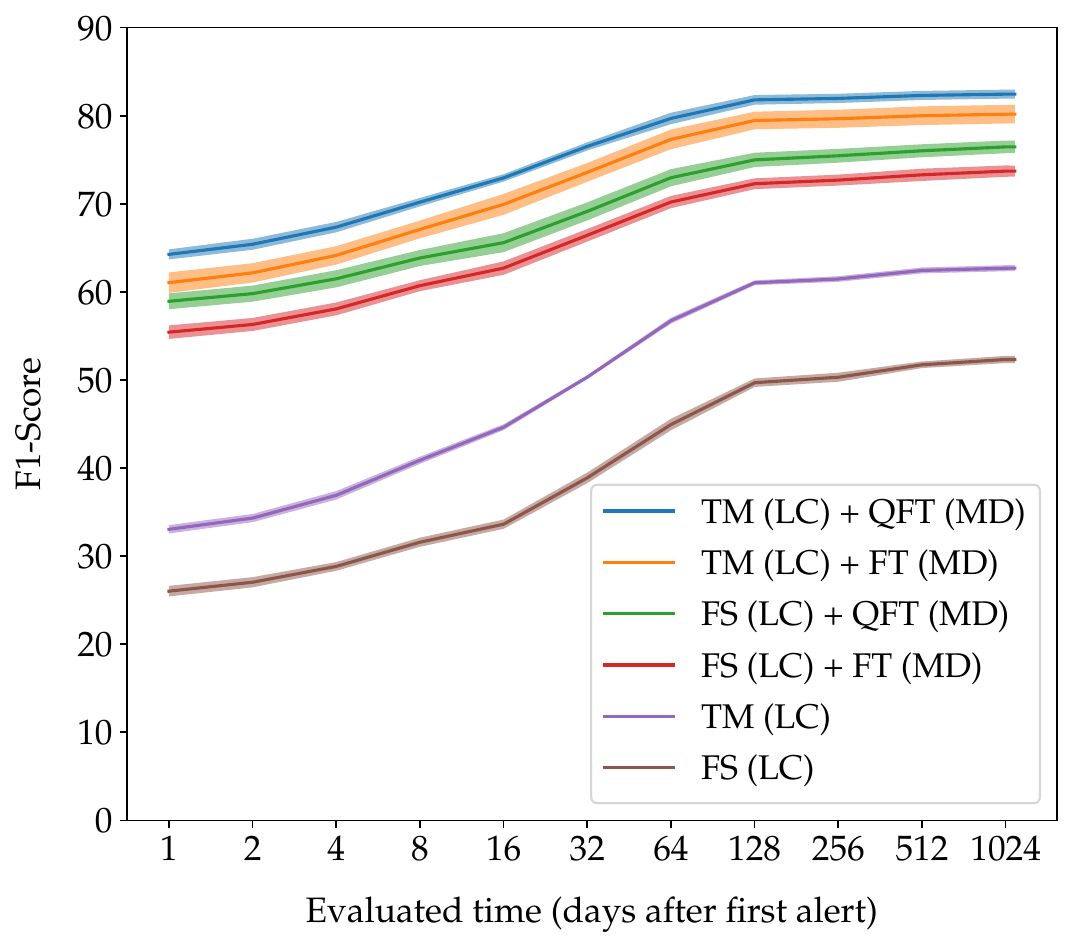}
      \caption{\label{fig:ablation} Ablation study in time. We compare combinations of our proposed time modulation (TM) and quantile feature tokenizer (QFT) as compared to a fixed sinusoid (FS) positional encoding (PE) and non-quantile feature tokenizer (FT). The combination of TM and QFT outperforms other combinations of PEs and tabular tokenizers independent of the time the models are evaluated.}
\end{figure}

Figure~\ref{fig:ablation} shows the performance of the models as a function of the days since the first alert for each of the combinations of PEs and tokenizers. This Figure shows the superiority of TM and QFT over the models that use a fixed sinusoidal PE or a non-quantile tabular data tokenizer. We conclude that there is a significant contribution of our proposed modification to the light curves and tabular data transformers for the classification of astronomical data streams from the ELAsTiCC challenge.

\subsection{Computational time performance}

Table \ref{tab:comp_time} shows the average computational time to predict the class of a single light-curve\footnote{Averages are estimated using the whole dataset and full-length light-curves.} with the selected ATAT variants and the RF-based baseline. The table also shows the average time per light-curve  to compute the complete set of engineered features. Note that only the RF-based baseline and ATAT (LC + MD + Features + MTA) require features to be computed. \rev{In light of the LSST emitting millions of alerts per night, evaluating inference times in batches is essential for assessing the practical feasibility of our approach. Inference times for ATAT on the GPU are shown for both a batch of 2,000 light-curves and one light-curve at a time (averaged over 20,000 batches of one light-curve each).}

From Table \ref{tab:comp_time}, we observe that the computational time required to perform inference with any of the models is negligible in comparison with the time required to compute features. This means that in total, the ATAT (LC + MD + MTA) variant, is orders of magnitude faster than the RF-based baseline and feature-based ATAT variants. This sets the LC + MD + MTA variant as a very interesting trade-off, reducing computational time in $99.75\%$ with only a $0.3\%$ decrease in F1-score. We note, however, that some classes are more affected when features are excluded, e.g., periodic variables. As future work we plan to explore which subset of features are more synergistic with the ATAT (LC + MD +  MTA) variant. The selection of the best trade-off may also need to be reevaluated as future surveys such as LSST are expected to incorporate some features (e.g., period) in the alert stream. 

\begin{table}[h]
\centering
\begin{tabular}{lc}
\toprule
Inference step    & Average time [s]\\
\midrule
{\bf CPU}\\
Feature computation       & $1.88 \cdot 10^{-1}$     \\
RF (MD + Features) &  $\mathbf{2.15} \cdot 10^{-4}$       \\
ATAT (LC + MD + MTA)       & $6.44 \cdot 10^{-3}$   \\
ATAT (LC + MD + Features + MTA)      & $1.34 \cdot 10^{-2}$     \\
\midrule
{\bf GPU (2,000 light-curves per batch)}\\
ATAT (LC + MD + MTA)       & $4.75 \cdot 10^{-4}$   \\
ATAT (LC + MD + Features + MTA)      & $8.29 \cdot 10^{-4}$     \\
\midrule
{\bf GPU (1 light-curve per batch)}\\
ATAT (LC + MD + MTA)       & $8.14 \cdot 10^{-3}$   \\
ATAT (LC + MD + Features + MTA)      & $8.45 \cdot 10^{-3}$     \\
\bottomrule
\end{tabular}
\vspace{.1cm}
\caption{\label{tab:comp_time} Average computational time  per light-curve in seconds required to perform the inference step for selected classification models. We used a single core of an AMD EPYC 7662 processor and a NVIDIA A100 GPU for these experiments.}
\end{table}

\section{Discussion and Conclusions}

We presented the models used by the ALeRCE team during the first round of ELAsTiCC. We introduced ATAT, a novel Deep Learning Transformer model that combines time series and tabular data information.
The proposed model was developed for the ELAsTiCC challenge that simulates an LSST-like stream, with the objective of testing  end-to-end alert stream pipelines. We were able to evaluate both classification and infrastructure performance metrics in the training set provided by ELAsTiCC. Our model was put into production within the ALeRCE broker in preparation for the real-time classification of the LSST alert stream.


Our results show that, using the ELAsTiCC dataset, ATAT outperforms a Balanced Hierarchical RF model similar to the current ALeRCE's light-curve classifier. This RF obtains a macro precision/recall/f1-score of 0.777/0.782/0.772, while ATAT achieves 0.841/0.827/0.825 when using light-curves, metadata and features calculated over the light-curves. Furthermore, if only the light-curves and metadata are considered for ATAT, we achieved values of 0.838/0.825/0.823 for the previous metrics, and, \rev{when performing inference in batches of 2,000 light-curves,} about 400 times faster inference times than with the RF. Importantly, our work suggests that it is possible to classify light-curves excluding human-engineered features with no significant loss in performance, and highlights the importance of including metadata information such as the properties of the host galaxy (e.g., \cite{2022AJ....164..195F}). These results may be improved for data-scarce classes by class-weighting and/or additional data-augmentation strategies \citep{Boone_2019}. We plan to explore these alternatives in the future.

The metrics presented in this work, e.g., in Table~\ref{tab:my_label} or Figure~\ref{fig:teval}, are representative of the dataset provided by ELAsTiCC to prepare machine learning models previous to the end-to-end challenge. The ELAsTiCC simulated data may not representative of the real LSST alert stream, and this  may result in different performance metrics than those reported in this work. In order to tackle these differences, we suggest the application of fine-tuning and domain adaptation techniques.

ATAT has proven to be competitive against feature-based tree ensembles in a large, complex and multi-class alerts light-curve classification setting, including very different variability classes. Transformer-based models represent a paradigm shift and we believe that more astronomical applications based on these models will be developed. Particularly, ATAT opens the door for more multi-modal applications, e.g., a third branch for stamps in Fig \ref{atat}.


\begin{acknowledgements}
The authors acknowledge support from the Chilean Ministry of Economy, Development, and Tourism's Millennium Science Initiative through grant ICN12\_009, awarded to the Millennium Institute of Astrophysics, and from the National Agency for Research and Development (ANID) grants: BASAL Center of Mathematical Modeling Grant PAI AFB-170001 (NA, IRJ, FF, JA, AMA, AA), BASAL projects ACE210002 and FB210003 (AB, MC), FONDEQUIP  Patagón supercomputer of Universidad Austral de Chile EQM180042 (NA, GCV, PH), FONDECYT Iniciación 11191130 (GCV), FONDECYT Regular 1200710 (FF), FONDECYT Regular 1211374 (PH), FONDECYT Regular 1220829 (PAE), and infrastructure fund QUIMAL190012. We acknowledge support from REUNA Chile. This work has been possible thanks to the use of Amazon Web Services (AWS) credits managed by the Data Observatory. ELAsTiCC is a project of the U.S. Department of Energy-supported Dark Energy Science Collaboration. ELAsTiCC used resources of the National Energy Research Scientific Computing Center (NERSC), a U.S. Department of Energy Office of Science User Facility located at Lawrence Berkeley National Laboratory. We are grateful to the team that created the public ELAsTiCC challenge: Gautham Narayan, Alex Gagliano, Alex Malz, Catarina Alves, Deep Chatterjee, Emille Ishida, Heather Kelly, John Franklin Crenshaw, Konstantin Malanchev, Laura Salo, Maria Vincenzi, Martine Lokken, Qifeng Cheng, Rahul Biswas, Renée Holžek, Rick Kessler, Robert Knop, Ved Shah Gautam.
\end{acknowledgements}

%
%
\bibliographystyle{aa}
\bibliography{bibliography.bib}

\begin{appendix} 
\section{Processed features details}
\label{processedfeat}
To extract color information from the difference light-curves, on each band we take the absolute value of the flux, compute the percentile 90 and save that value. Following the order ugrizY, we take the value of  the percentile 90 previously saved for one band and divide it by the value of the next band. To avoid dividing by zero, we add 1 to the denominator.

Most differences in the features used are related to extracting information from supernova-like light-curves. This is the list of supernova features, which are computed for each band:

\begin{itemize}
    \item \verb+positive_fraction+: fraction of observations with a positive flux value.
    
    \item \verb+dflux_first_det_band+: difference between the flux of the first detection (in any band) and the last non-detection (in the same selected band) just before the first detection.
    
    \item \verb+dflux_non_det_band+: same as \verb+dflux_first_det_band+, but instead of using the last non-detection before the first detection, we take all the non-detections before the first detection and compute the median. Later, this median is subtracted from the flux of the first detection (in any band).
    
    \item \verb+last_flux_before_band+: flux of the last non-detection (in the selected band) before the first detection (in any band).
    
    \item \verb+max_flux_before_band+: maximum flux of the non-detections (in the selected band) before the first detection (in any band).
    
    \item \verb+max_flux_after_band+: maximum flux of the non-detections (in the selected band) after the first detection (in any band).
    
    \item \verb+median_flux_before_band+: median flux of the non-detections (in the selected band) before the first detection (in any band).
    
    \item \verb+median_flux_after_band+: median flux of the non-detections (in the selected band) after the first detection (in any band).
    
    \item \verb+n_non_det_before_band+: number of non-detections (in the selected band) before the first detection (in any band).
    
    \item \verb+n_non_det_after_band+: number of non-detections (in the selected band) after the first detection (in any band).
\end{itemize}

As we were not sure if the ELAsTiCC stream would indicate if the observations were alerts or forced photometry (i.e. if the signal was strong enough compared with the noise), for the supernova features we considered an observation as a detection if the absolute value of the difference flux was at least 3 times larger than the observation error.

With respect to the Supernova Parametric Model \citep[SPM, ][]{Sanchez-Saez_2021}, one SPM model per band was fitted to the data, but the optimization was done simultaneously and penalizing the dispersion between the parameters on different bands. The extra term added to the cost function is 
\begin{equation}
    \left \langle 
        \begin{pmatrix}
        \widehat{Var}(A) + 1\\ 
        \widehat{Var}(t_0) + 0.05\\ 
        \widehat{Var}(\gamma) + 0.05\\
        \widehat{Var}(\beta) + 0.005\\
        \widehat{Var}(t_{rise}) + 0.05\\
        \widehat{Var}(t_{fall}) + 0.05\\
        \end{pmatrix},
        \begin{pmatrix}
        0.0\\
        1.0\\
        0.1\\
        20.0\\
        0.7\\
        0.01
        \end{pmatrix}
     \right \rangle
\end{equation}
where the variances are estimated over the different bands and the coefficients were found experimentally. The original SPM code was modified to avoid numerical instabilities. To speed up the optimization, the gradient of the cost function is computed using the JAX library \footnote{\url{http://github.com/google/jax}}. The initial guess and the boundaries for the parameter optimization were tuned for the range of values in the ELAsTiCC dataset.

\end{appendix}

\end{document}